\documentclass[pre,reprint,twocolumn,showpacs,superscriptaddress,preprintnumbers]{revtex4-1}
\usepackage{amsmath}
\usepackage{graphicx}
\usepackage{graphics}
\usepackage{epstopdf}
\usepackage{bm}
\newcommand{\Om}{\Omega\left[\ro\right]}

\newcommand{\intdr}{\int d{\bf r}}
\newcommand{\intdrp}{\int d{\bf r'}}

\newcommand{\ro}{\rho\left({\bf r}\right)}
\newcommand{\rop}{\rho\left({\bf r}'\right)}

\begin{document}
\title{Geometry-induced phase transition in fluids: capillary prewetting.}
\author{Petr Yatsyshin}
\affiliation{Department of Chemical Engineering, Imperial College London,
London SW7 2AZ, United Kingdom}
\author{Nikos Savva}
\affiliation{Department of Chemical Engineering, Imperial College London,
London SW7 2AZ, United Kingdom} \affiliation{Cardiff School of Mathematics,
Cardiff University, Cardiff, CF24 4AG, United Kingdom}
\author{Serafim Kalliadasis}
\affiliation{Department of Chemical Engineering, Imperial College London,
London SW7 2AZ, United Kingdom}
\date{\today}


\begin{abstract}
We report a new first-order phase transition preceding capillary condensation
and corresponding to the discontinuous formation of a curved liquid meniscus.
Using a mean-field microscopic approach based on the density functional
theory we compute the complete phase diagram of a prototypical
two-dimensional system exhibiting capillary condensation, namely that of a
fluid with long-ranged dispersion intermolecular forces which is spatially
confined by a substrate forming a semi-infinite rectangular pore exerting
long-ranged dispersion forces on the fluid.  In the $T$--$\mu$ plane the
phase line of the new transition is tangential to the capillary condensation
line at the \emph{capillary wetting temperature}, $T_{\text{cw}}$. The surface phase behavior of the system maps to planar wetting with the phase line of the new transition, termed \emph{capillary prewetting}, mapping to the planar prewetting line.
If capillary condensation is approached isothermally with $T>T_{\text{cw}}$,
the meniscus forms at the capping wall and unbinds continuously, making capillary
condensation a second-order phenomenon. We compute the
corresponding critical exponent for the divergence of adsorption.
\end{abstract}


\maketitle


Capillary wetting is crucial in a wide variety of natural processes and
technological applications. From the wetting properties of plant
leaves~\cite{GouLiuPlantSci07} to building of nanoreactors~\cite{Dum09} and
design of ``lab-on-chip" devices~\cite{BerardinoDietrich2012AndSquires05}. It
also provides one of the most striking manifestations of the attractive
intermolecular forces governing the behavior of matter. For microscopic
systems, concepts such as surface tensions and contact angles become quite
limited. One must account for the molecular structure of the fluid, since
most non-trivial effects in confinement are caused by the interplay of
different length scales corresponding to particle sizes, ranges of
intermolecular potentials and dimensions of confinement. These parameters act
as independent thermodynamic fields, leading, according to the Gibbs rule, to
a rich surface phase behavior and metastability caused by the interplay of
different wetting mechanisms~\cite{Evans90,HendersonBook92}. The main
theoretical approaches to confined fluids are mean field Van der Waals loops,
renormalization group theory, and effective Hamiltonians. While the latter
two have been successfully applied to establishing the universality of some
surface critical phenomena,
e.g.~\cite{MilchevPRL03AndSartoriParryJPhysCondensMatt02,TasDietPRL06}, far
less attention has been given to the molecular structure of an adsorbate.
Here we adopt a fully microscopic approach based on density functional (DF)
theory leading to a molecular model for the fluid, undertaking a systematic
investigation of wetting in a semi-infinite rectangular pore. The scenario of
drying by a fluid with short-ranged forces and zero contact angle has been
considered within a DF approach in Ref.~\onlinecite{Roth11}, but without
fully exploring the phase diagram. Here we obtain detailed information on
density profiles and menisci shapes, adsorption isotherms, phase diagrams and
critical behavior. Noteworthy is also that our approach is not restricted to
complete wetting scenarios as in previous works, e.g.,
Refs.~\onlinecite{TasDietPRL06,DarbellayYeomans92}.

Consider a long-range interacting fluid confined to a rectangular pore of
width $H$ and length $L$ by a substrate whose walls exert long-ranged dispersive forces on the fluid.  The system is assumed infinite in the direction normal to the $H$--$L$ plane, where it is also connected to a thermostat fixing the values of the chemical potential $\mu$ and temperature $T$. Under these assumptions the density profile of the fluid inside the pore is two dimensional (2D): $\rho\left({\bf r}\right)\equiv\rho\left(x,y\right)$. The free energy of the pairwise interacting fluid in an external field $V\left(x,y\right)$, formed by the substrate walls, is a unique functional of the one-body density, which in the grand canonical ensemble is given by
\begin{equation}
    \label{Om}
    \Om=F\left[\rho(\mathbf{r})\right]-\mu\intdr\ro.
\end{equation}

An approximation for the canonical free energy functional $F$ follows from the perturbation around a hard sphere fluid in powers of the attractive
potential $\varphi_\text{attr}$ up to the first order~\cite{Evans}:
\begin{align}
    \label{F}
    F\left[\ro\right] &= \intdr \left(f_{\text{id}}\left(\rho\right)+ \rho\psi\left(\rho\right)+\rho V\right)\notag\\
    &+\frac{1}{2}\intdr\intdrp\ro\rop\varphi_\text{attr}\left(\left|{\bf r}-{\bf r}'\right|\right),
\end{align}
where $f_{\text{id}}\left(\rho\right)=k_{\text{B}}T\int d{\bf
r}\rho\left(\ln\lambda^3\rho-1\right)$ is the ideal free energy,
$k_{\text{B}}$ is the Boltzman's constant and $\lambda$ -- the thermal
wavelength, and $\psi\left(\rho\right)$ is the configurational part of the
free energy. Using the Carnahan and Starling equation of state for the gas of
hard spheres with diameter $\sigma$, it can be approximated as ~\cite{CS}
\begin{equation}
    \label{psi}
    \psi\left(\rho\right)= k_{\text{B}}T\frac{\eta\left(4-3\eta\right)}{\left(1-\eta\right)^2},\quad\eta=\pi\sigma^3\rho/6.
\end{equation}
The long-ranged fluid-fluid forces can be described following the Barker and Henderson prescription~\cite{BH}
\begin{equation}
    \varphi_{\mbox{attr}}\left(r\right) = \left\{
    \begin{array}{lr}
        0, &r\leq \sigma\\
        4\varepsilon\left[\left(\dfrac{\sigma}{r}\right)^{12}-\left(\dfrac{\sigma}{r}\right)^{6}\right], &r>\sigma
    \end{array}
    \right.,
    \label{LJ}
\end{equation}
with $\varepsilon$ being the strength parameter, which we will be using as the unit of energy.

Finally, for simplicity and without loss of generality, we model the
substrate by
\begin{equation}
    \label{wall}
V\left(x,y\right)=V_{0}^{\text{B}}\left(y\right)+V_{0}^{\text{L}}\left(x\right)+V_{0}^{\text{T}}\left(H-y\right),
\end{equation}
 ($L\to\infty$) where the superscript stands for bottom, left and top walls respectively, and each
\begin{equation}
    \label{dwall}
    V_0^{\left(\text{i}\right)}\left(z\right)=E_0^{\left(\text{i}\right)}\left(-\frac{1}{6}\left(\frac{\sigma_{0}}{z+z_0}\right)^3+\frac{1}{45}\left(\frac{\sigma_{0}}{z+z_0}\right)^9\right)
\end{equation}
is the 3-9 potential of an isolated planar wall characterized by the
parameters of strength $\varepsilon_{\text{0}}$, range $\sigma_{0}$ and
low-$z$ cut-off $z_0$;
$E_{0}^{\left(\text{i}\right)}=4\pi\varepsilon_{\text{0}}^{\left(\text{i}\right)}\rho_0\sigma_{\text{0}}^3$
\cite{Yatsyshin2012}. Throughout this letter we fix the value of the
capillary width $H=30\sigma$ and the wall parameters: $\rho_0=1\sigma^{-3}$,
$\sigma_0=2\sigma$, $z_0=5\sigma$. Presented examples differ in the values of
$\varepsilon_{\text{0}}^{\left(\text{i}\right)}$, which will be provided
along with the wetting temperatures $T_{\text{w}}^{\left(\text{i}\right)}$
and macroscopic contact angles at bulk coexistence
$\Theta^{\left(\text{i}\right)}$ for each $V_{0}^{\left(\text{i}\right)}$ in
Eq.~\eqref{wall}. For complete prewetting lines see Supplemental Material:
``1D Wetting".

The model for substrate potential \eqref{wall} was chosen for computational
convenience, as well as to present several non-symmetric examples in a
straightforward way. Alternatively, considering a symmetric case, one could
integrate the pairwise Lennard-Jones potential over the exterior of the
capillary. However, this would not alter the forthcoming discussion, since
the physics of the external field comes down to the effect of long-ranged
pairwise London forces acting between the fluid and the substrate particles,
and is inherent in both prescriptions. Moreover, the presence of the
discussed phenomena is not restricted by a particular model of the substrate,
so long as it arises from the dispersive pairwise forces.

The equilibrium configurations $\rho\left(x,y\right)$ minimize \eqref{Om} and
are computed numerically using a 2D extension of the method developed in
Ref.~\onlinecite{Yatsyshin2012}. The 2D scheme was verified by obtaining 1D
profiles with it, as well as by checking the agreement with contact sum rules
\cite{HendersonBook92}. Isotherms and phase diagrams were obtained using
arc-length continuation.

The DF \eqref{F} is known to capture the physics of fluids at vapor- and
liquid-like densities. Using a different expression for the substrate
potential (e.g., an \emph{ab initio} potential~\cite{Ust05Marsh96AndChiz98}),
as well as a different version of the DF (e.g., accounting for the repulsive
effects of molecular packing non-locally in $\rho\left({\bf r}\right)$
through an improved term $\psi\left(\ro\right)$ in
\eqref{F}~\cite{WhiteBear}), would only induce minor quantitative changes but
not add to the understanding of the physical phenomena described here and
adequately captured by \eqref{F}~\cite{HendersonBook92}.
\begin{figure}[h]
    \includegraphics[width=1\linewidth]{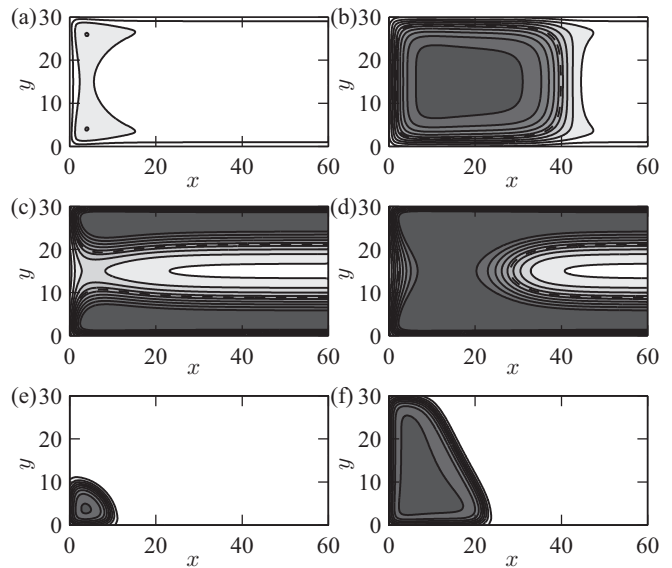}
    \caption{Density profiles in non-symmetrical capped capillaries coexisting during capillary prewetting ($\Delta\mu_{\text{cpw}}$), which precedes capillary condensation ($\Delta\mu_{\text{c}}$). The data is rescaled between capillary $\rho_{\text{vap}}$ (white) and $\rho_{\text{liq}}$ (black); dashed line marks the liquid-vapor interface.
    \\
    (a) and (b): $T=.95T_{\text{c}}$, $\Delta\mu_{\text{c}}=-8.4\cdot10^{-3}\varepsilon$, $\Delta\mu_{\text{cpw}}=-9\cdot10^{-3}\varepsilon$, $\varepsilon_{\text{0}}^{\text{B, L, T}}=.3\varepsilon$, $T_{\text{w}}^{\text{B, L, T}}=.96T_{\text{c}}$, $\Theta^{\text{B, L, T}}=31.2^{\circ}$;
    (c) and (d): $T=.96T_{\text{c}}$, $\Delta\mu_{\text{c}}=-.058\varepsilon$, $\Delta\mu_{\text{cpw}}=-.06\varepsilon$, $\varepsilon_{\text{0}}^{\text{B, T}}=.8\varepsilon$, $\varepsilon_{\text{0}}^{\text{L}}=.3\varepsilon$, $T_{\text{w}}^{\text{B, T}}=.71T_{\text{c}}$, $T_{\text{w}}^{\text{L}}=.96T_{\text{c}}$, $\Theta^{\text{B, L, T}}=0$;
    (e) and (f): $T=.72T_{\text{c}}$, $\Delta\mu_{\text{c}}=-.018\varepsilon$, $\Delta\mu_{\text{cpw}}=-.032\varepsilon$, $\varepsilon_{\text{0}}^{\text{B, L}}=.8\varepsilon$, $\varepsilon_{\text{0}}^{\text{T}}=.3\varepsilon$, $T_{\text{w}}^{\text{B, L}}=.71T_{\text{c}}$, $T_{\text{w}}^{\text{T}}=.96T_{\text{c}}$, $\Theta^{\text{B, L}}=0$, $\Theta^{\text{T}}=109.25^{\circ}$.\label{FigOne}}
\end{figure}

As $x\to\infty$, $\rho\left(x,y\right)\to\rho_{\text{1D}}\left(y\right)$, with $\rho_{\text{1D}}\left(y\right)$ being the density distribution inside a slit pore of width $H$. The requirement to form two liquid-wall interfaces leads to the Kelvin shift of the bulk coexistence curve given (in the symmetric case $V_{0}^{\text{B}}\equiv V_{0}^{\text{T}}$) by~\cite{SaamJLowTempPhys09}
\begin{equation}
P_{\text{vap}}-P_{\text{liq}}=\dfrac{2\sigma_{\text{lv}}\cos\Theta}{H}+\text{h.o.t.},
\label{Kelvin}
\end{equation}
where $\sigma_{\text{lv}}$ is the liquid-vapor surface tension, $\Theta$ is
the contact angle, and the pressures $P_{\text{vap}}$ and $P_{\text{liq}}$
refer to capillary-vapor and -liquid phases, which, due to \eqref{Kelvin},
can coexist inside the pore. These phases transform through a discontinuous first-order transition known as capillary condensation (CC)~\cite{Evans90,HendersonBook92}, which is essentially a shifted bulk liquid-vapor coexistence.

The phase behavior of a capped capillary is closely linked to that of the associated slit pore. For example, when CC is approached isothermally, the capped capillary can undergo a continuous second-order transition analogous to complete wetting of planar walls \cite{Parry07}. It corresponds to the gradual unbinding of the liquid meniscus from the capping wall and filling the capillary. Can the analogy with 1D wetting be extended beyond that observation?

Our computations indicate the presence of a discontinuous first-order
transition corresponding to the formation of the liquid meniscus at a finite
distance from the capping wall, as the fluid's chemical potential (pressure)
is increased isothermally towards CC. Depending on the ``wall set", Eqs. \eqref{wall}, \eqref{dwall}, the topology of coexisting profiles can vary quite significantly, as shown in Fig.~\ref{FigOne}; also see Supplemental
Material, movies 1--3. For illustration purposes, the 2D
data is rescaled between capillary $\rho_{\text{vap}}$ (white) and
$\rho_{\text{liq}}$ (black). The interface (dashed curves) is defined as
$\left(\rho_{\text{liq}}+\rho_{\text{vap}}\right)/2$ and the temperature is
measured in units of the bulk critical temperature $T_{\text{c}}$.

\begin{figure}[h]
\includegraphics{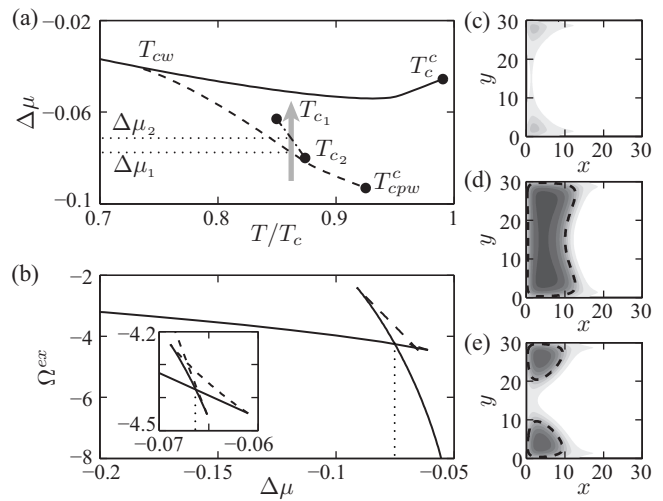}
    \caption{(a): Phase diagram of the symmetric capped capillary with $\varepsilon_{\text{0}}^{\text{B, L, T}}=.7\varepsilon$ and $T_{\text{w}}^{\text{B, L, T}}=.76T_{\text{c}}$.
    Full line: $\Delta\mu_{\text{c}}\left(T\right)$ -- capillary condensation line of a slit pore associated with the capillary, terminating at the capillary critical temperature $T_{\text{c}}^{\text{c}}=.98T_{\text{c}}$.
    Dashed line between $T_{\text{cw}}=.735T_{\text{c}}$ ($\Theta^{\text{B, L, T}}=24.8^{\circ}$) and $T_{\text{cpw}}^{\text{c}}=.93T_{\text{c}}$ running tangent to $\Delta\mu_{\text{c}}\left(T\right)$: $\Delta\mu_{\text{cpw}}\left(T\right)$ --
    capillary prewetting line (see text).
    Dashed line between $T_{{\text{c}}_1}=.85T_{\text{c}}$ and $T_{{\text{c}}_2}=.875T_{\text{c}}$: $\Delta\mu_{\text{c}_{12}}\left(T\right)$ --
    corner prefilling line (here is metastable).
    (b): isotherm of the route designated in (a) by an arrow at $T_0=.855T_{\text{c}}$  ($\Theta^{\text{B, L, T}}=0$); inset zooms on corner prefilling transition (at $\Delta\mu_{\text{2}}$), vertical dotted lines correspond to $\Delta\mu_{\text{1}}=-.075\varepsilon$ and $\Delta\mu_{\text{2}}=-.067\varepsilon$ --
    intersections of the route with the phase lines of capillary prewetting and corner prefilling transitions. (c)-(e): typical coexisting density profiles: (c) - vapor, (d) - liquid slab with meniscus, (e) - droplets in the corners.\label{FigTwo}}
\end{figure}

For simplicity we focus on the symmetric case of a capped capillary formed by
identical walls, Eqs. \eqref{wall} and \eqref{dwall}, with the aim of
investigating its surface phase behavior, Fig.~\ref{FigTwo}. Obtaining the
phase diagram of the system is non-trivial and consists of several steps.
First, we compute the CC line of the associated 1D slit pore, to which the
system reduces as $x\to\infty$ (full line,
$\Delta\mu_{\text{c}}\left(T\right)$ on Fig.~\ref{FigTwo}(a), where
$\Delta\mu$ is counted from the bulk liquid-vapor coexistence). Next, we
compute an isotherm at a more or less arbitrary temperature $T_0$ below the
capillary critical temperature $T_{\text{c}}^{\text{c}}$. Figure
\ref{FigTwo}(b) shows $\Omega^{\text{ex}}\left(\mu\right)$ for
$T_0=.855T_{\text{c}}$, with the excess-over-slit grand free energy defined
as
$\Omega^{\text{ex}}=\Omega\left[\rho\left(x,y\right)\right]-\Omega\left[\rho_{\text{1D}}\left(y\right)\right]$.
Two typical Van-der-Waals hysteresis loops reveal the presence of first-order
phase transitions: a stable one at $\Delta\mu_{1}=-.75\varepsilon$ and a
metastable one at $\Delta\mu_{2}=-.67\varepsilon$ (see inset on
Fig.~\ref{FigTwo}(b)). The stable transition corresponds to coexisting
configurations of capillary vapor, Fig.~\ref{FigTwo}(c), and capillary-liquid
slab, Fig.~\ref{FigTwo}(d). For the metastable transition the coexisting
configurations are those of capillary-vapor (indistinguishable from
Fig.~\ref{FigTwo}(c)) and two capillary-liquid drops in the corners,
Fig.~\ref{FigTwo}(e). Now, applying arc-length continuation in the manner
described in ~\cite{Yatsyshin2012,FrinkSalinger1},
we calculate the complete phase lines of the above transitions. The results are plotted on Fig.~\ref{FigTwo}(a) by a dashed line for the stable transition and a dashed-dotted line for the metastable one.

Finally, we ensure that the phase behavior of the system is fully scrutinised
and there are no other transitions by calculating a set of isotherms for a
broad range of temperatures and identifying the phase transitions revealed by
those isotherms with the phase lines obtained above. Each isotherm spans a
broad range of values of $\Delta\mu$ starting far from CC, where an
equilibrium configuration is that of capillary-vapor (similar to
Fig.~\ref{FigTwo}(c)), and ending with $\Delta\mu$ very close to CC, where
the capillary is filled. Moreover, after investigating various sets of
parameters (non-symmetric capillaries, strengths and ranges of wall
potentials, capillary widths, etc.), we can confirm that the phase diagram
and the isotherm presented on Figs. \ref{FigTwo}(a) and \ref{FigTwo}(b) are
quite typical and thus capture the essential physics of the system.

Consider first the $\Delta\mu_{\text{cpw}}\left(T\right)$-line (dashed line
on Fig.~\ref{FigTwo}(a)) forming the locus of discontinuous transitions to
fluid configurations of the type shown on Fig.~\ref{FigTwo}(d). It runs
tangent to the CC line $\Delta\mu_{\text{c}}\left(T\right)$ (full line on
Fig.~\ref{FigTwo}(a)) at the point $T_{\text{cw}}$ and ends at
$T_{\text{cpw}}^{\text{c}}$. For thermodynamic routes with constant $T$
crossing the $\Delta\mu_{\text{cpw}}\left(T\right)$-line vertically we
observe a discontinuous formation of the liquid meniscus at a finite distance to the capping wall, which unbinds to infinity as CC is approached ($\Delta\mu\to\Delta\mu_{\text{c}}$). Above the critical temperature $T_{\text{cpw}}$ the liquid meniscus at the capping wall is being formed continuously.

For values of $T$ 
chosen closer to $T_{\text{cw}}$ from above, the meniscus forms at larger
distances from the capping wall. In the limit $T=T_{\text{cw}}$ and
$\Delta\mu\to\Delta\mu_{\text{c}}\left(T_{\text{cw}}\right)$ there is a
discontinuous jump to CC: the whole capillary becomes filled with
capillary-liquid.

For isothermal routes approaching $\Delta\mu_{\text{c}}\left(T\right)$ at $T<T_{\text{cw}}$, the capped capillary remains in capillary-vapor (Fig.~\ref{FigTwo}(c)), until at $\Delta\mu=\Delta\mu_{\text{c}}\left(T\right)$ the CC, occurring discontinuously in the spectator phase ($x\to\infty$), drives it into capillary-liquid. The CC in that case occurs ``from capillary bulk'', in contrast to the case of $T>T_{\text{cw}}$, when it happens ``from the surfaces'' of the capped capillary.
This phenomenon is analogous to partial wetting in 1D, when for
temperatures less than the wetting temperature of the planar wall the fluid
does not adsorb an infinite liquid layer at coexistence ($\Theta>0^{\circ}$).

In fact, all the described phase behavior of the capped capillary is in complete analogy with the wetting phenomenology of a single planar wall immersed in vapor, when the wetting transition is first-order~\cite{SaamJLowTempPhys09}. The CC line
$\Delta\mu_{\text{c}}\left(T\right)$ maps to the bulk coexistence line
($\Delta\mu=0$),

\begin{figure}[h]
\includegraphics{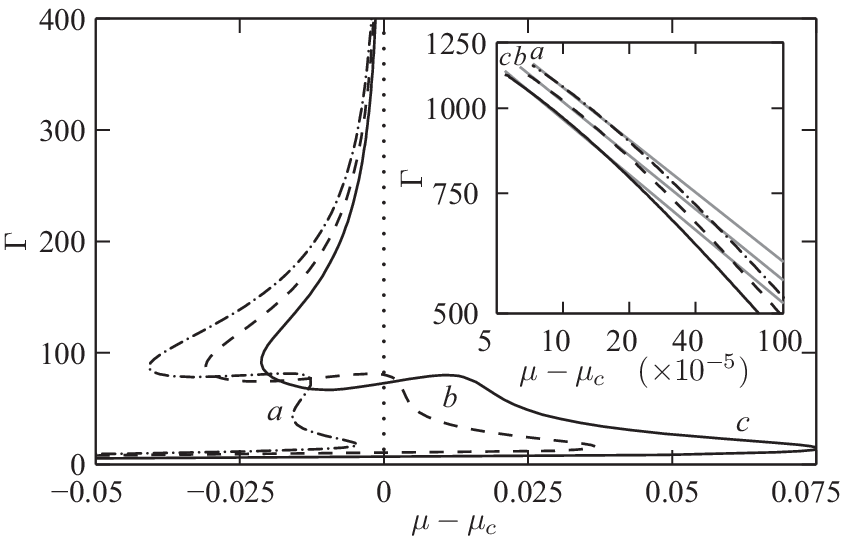}
    \caption{Adsorption isotherms at $T=.85T_{\text{c}}$ of symmetric capped  capillaries, $\Delta\mu\to\Delta\mu_{\text{c}}$. Inset: Estimated critical exponents (black) with a guide to eye (grey): $\Gamma\sim\left(\mu-\mu_{\text{c}}\right)^{-1/4}$.
    \\
     \emph{a}: $\varepsilon_{\text{0}}^{\text{B, L, T}}=.7\varepsilon$, $T_{\text{w}}^{\text{B, L, T}}=.76T_{\text{c}}$, $\Theta^{\text{B, L, T}}=0$, $\Gamma\sim\left(\mu-\mu_{\text{c}}\right)^{-.256}$,
    \emph{b}:  $\varepsilon_{\text{0}}^{\text{B, L, T}}=.6\varepsilon$, $T_{\text{w}}^{\text{B, L, T}}=.82T_{\text{c}}$, $\Theta^{\text{B, L, T}}=0$,  $\Gamma\sim\left(\mu-\mu_{\text{c}}\right)^{-.262}$,
    \emph{c}:  $\varepsilon_{\text{0}}^{\text{B, L, T}}=.5\varepsilon$, $T_{\text{w}}^{\text{B, L, T}}=.87T_{\text{c}}$, $\Theta^{\text{B, L, T}}=26.8^{\circ}$,  $\Gamma\sim\left(\mu-\mu_{\text{c}}\right)^{-.269}$. \label{FigThree}}
\end{figure}
\noindent and indicates the limiting values of $\Delta\mu$ at which
the spectator fluid phase is expected to transform. Likewise, the
$\Delta\mu_{\text{cpw}}\left(T\right)$-line maps onto the prewetting line. We therefore refer to $T_{\text{cw}}$ as the \emph{capillary wetting
temperature}, to the phase transition at $T_{\text{cw}}$ and $\Delta\mu_{\text{cpw}}\left(T_{\text{cw}}\right)$ as \emph{first-order capillary wetting}, and to the transitions corresponding to $\Delta\mu_{\text{cpw}}\left(T\right)$-line -- as \emph{capillary
prewetting}. Obviously, if $H\to\infty$, $T_{\text{cw}}\to T_{\text{w}}$, the
wetting temperature of the capping wall.

An interesting question arises whether, for some specific sets of substrate
parameters, the transition to CC at $T_{\text{cw}}$ can be continuous and not
first-order. Following the observed analogy with planar wetting, such
scenario should be possible and a tricritical line separating the first- and
second- order CC can exist in the parameter space, even though our numerical
experiments did not exhibit a second-order transition to capillary filling.
By the same token, second-order capillary filling can be expected for fluids
with short-ranged interactions.

Consider the dashed-dotted phase line on Fig.~\ref{FigTwo}(a),
$\Delta\mu_{\text{c}_{12}}\left(T\right)$, which corresponds to a
discontinuous formation of capillary-liquid drops in the corners, when
crossed isothermally (Fig.~\ref{FigTwo}(e)). This transition is a case of
corner-prefilling investigated in, e.g., Ref.~\onlinecite{RejDietNapPRE99}.
In the present case its phase line is metastable, but still affects the
topology of the density profiles coexisting during the capillary prewetting
transition. For isothermal routes crossing
$\Delta\mu_{\text{cpw}}\left(T\right)$ at $T<T_{{\text{c}}_2}$ capillary
prewetting takes place between a vapor phase (e.g., Fig.~\ref{FigTwo}(c)) and
a liquid slab phase (e.g., Fig.~\ref{FigTwo}(d)). For $T>T_{{\text{c}}_2}$
the formation of droplets in the corners is continuous as
$\Delta\mu\to\Delta\mu_{\text{cpw}}\left(T\right)$, and the capillary
prewetting transition then takes place between the configurations similar to
Fig.~\ref{FigTwo}(e) and Fig.~\ref{FigTwo}(d).

We note that the phase-line $\Delta\mu_{\text{c}_{12}}\left(T\right)$ is of
lesser importance for this work, being a particular case of
corner-prefilling~\cite{RejDietNapPRE99}; it is not related to CC (unlike the
phase line of capillary prewetting, which defines $T_{\text{cw}}$). For wider
capillaries, whose corners are further apart, the corner prefilling
transition will become stable, a triple point will exist, with the coexisting
fluid configurations of types shown in Figs.~\ref{FigTwo}(c)- (e).

Consider now the \emph{complete filling} of the capillary as $\Delta\mu\to\Delta\mu_{\text{c}}$ isothermally, above $T_{\text{cw}}$. Using an effective
interfacial Hamiltonian model, Parry~\emph{et al.}~\cite{Parry07} have
obtained the critical exponent for the diverging length of the liquid slab. Our
fully microscopic approach does not impose the existence of an interface; rather it is
obtained as a consequence of the non-uniformity of
the fluid caused by inter-molecular attractions. We consider the physical order
parameter of the system as $\Delta\mu\to\Delta\mu_{\text{c}}$, the conjugate thermodynamic variable, i.e. the adsorption:
\begin{equation}
\Gamma=\int d{\bf r}\left(\rho\left(x,y\right)-\rho_{\text{1D}}\left(y\right)\right),
\end{equation}
where integration is carried out over the entire volume of the capillary.

To properly capture the asymptote of the smooth decay of
$\rho\left(x,y\right)$ into the ``capillary bulk'', which ultimately determines
the critical exponent, we used a non-uniform grid, sufficiently dense
in the interval $x\leq100\sigma$ to resolve the liquid-vapor interface,
while imposing the boundary condition of contact with the slit pore at
$x\sim10^{4}\sigma$. The calculations were carried out for a range of temperatures, wall
potentials and capillary sizes without truncating the tails of potentials
\eqref{LJ} and \eqref{wall}. Figure \ref{FigThree} summarizes our results on
the criticality of CC in a capped capillary, which agree well with the
analytical work in Ref.~\onlinecite{Parry07}, where it was found that
$\Gamma\sim\left(\mu-\mu_{\text{c}}\right)^{1/4}$.

We believe that our results will motivate further studies on the role of
confinement in fluids. The new phase transition described here is potentially
an experimentally realizable scenario resulting from the interplay between
intra-fluid and fluid-substrate potentials. On the theoretical front, a
second-order transition to CC analogous to critical planar wetting may be
possible at $T_{\text{cw}}$, prompting an effective Hamiltonian treatment to
look for the tricritical line. Moreover, seeking a universality of capillary
prewetting among systems undergoing CC is well worth investigating, e.g. it
could motivate a renormalization group analysis, allowing also to explore the
role of fluctuations. Finally, the presence of metastability promises a rich
dynamic behaviour.
\begin{acknowledgments}
We are grateful to the anonymous referee for valuable comments and
suggestions. We acknowledge financial support from the European Research
Council via Advanced Grant No.\ 247031 and the European Framework 7 via Grant
No.\ 214919 (Multiflow).
\end{acknowledgments}


\end{document}